# Epistemic Reflections on AI Answering Our Questions: Overwatch, Erudite, Logician, Interlocutor

Johan F. Hoorn[1,*] and Ella-Jenna Oosterglorenwoud[2]

[1] The Hong Kong Polytechnic University, School of Design and Dept. of Computing
[2] Eindhoven University of Technology (TU/e), Industrial Engineering and Innovation Sciences

* Corresponding author: johan.f.hoorn [ad] polyu [dot] edu [dot] hk

## Abstract

Currently, there is a trend for the wider public to rely on Large Language Models (LLMs) for a variety of purposes (e.g., financial or legal consultation, medical and mental support, cf. Chatterji et al., 2025), often accepting the advice provided without necessarily seeking logical verification or empirical validation. While one might be fortunate enough to encounter a model with a particularly solid 'ground truth' or with auxiliary logic-symbolic reasoning capabilities, it remains a somewhat uncertain endeavour. All too often, the output of whatever system at hand is simply taken at face value, without further question. We argue that such careless reliance on AI to answer our questions and to judge our output is a violation of Grice's Maxim of Quality as well as a violation of Lemoine's legal Maxim of Innocence. A low-sensitivity plagiarism scanner may produce a Type II error by failing to detect difference (the null hypothesis wrongly maintained). The fallacy of *affirming the consequent* occurs when the failure to detect difference is then interpreted as evidence of equivalence or demonstration of AI authorship. If the test is specified so that 'AI-generated' is effectively treated as the default (null) hypothesis, then a finding of 'no difference from AI' is taken as support for that null. Such a mis-specified test results in students being treated as guilty (AI/plagiarism) unless suspects can generate sufficient detectable difference from AI output, which yields false accusations under a fair null hypothesis (that the student wrote the work). To avoid LLMs becoming a sorcerer's apprentice, knowledge is required about which inference systems are or should become integrated for an LLM to become a trustworthy sparring partner. We end on a wider perspective where the formalisation of the observer effect shows that uncertainty, classification, and interpretation are already shaped by the human or artificial agency's belief system, affective state, and tolerance for ambiguity, rather than arising only at the stage of LLM output or formal verification. We therefore do not wish to replace neuro-symbolic or verifier-centric AI, but reframe them within a broader account of trustworthy AI in which observer assumptions about data and data classes, as well as symbolic warrant, must be made explicit.

## 1. Introduction

For starters, we should perhaps acknowledge that, with generative and agentic AI currently evolving at warp speed, even the most contemporary insights may have a relatively short shelf-life. Nevertheless, human beings do not evolve at the same pace as AI and so with the advent of Deepfakes, Dall-e, and the GPT series, a tendency towards complacency can be observed (Lembke, 2021), individuals relying uncritically on neural networks (NNs) for content generation. At first glance, the output generated for artistic or entertainment purposes may appear innocuous (Lembke,

Humphreys, & Newmark, 2016). However, humans are more susceptible to influence than we might assume (e.g., Moore et al., 2026; Sharma et al., 2023). Although we may subconsciously recognise what is artificial, our cognitive processes tend to diminish the distinction between genuine and fabricated content, leading us to internalise the information as part of our personal perception. Memory, which is notoriously difficult to alter with precision, can be subtly modified, subliminally edited, without our conscious awareness: LLMs seem to be capable of shaping someone's personality (Li et al., 2026; Sharma et al., 2023).

The allure of a NN lies in its ability to produce outputs that appear relevant and coherent to the untrained observer (Chatterji et al., 2025). However, because it is nearly impossible to trace and identify the origins of these outputs (i.e., the ground 'truth' of the training set), the basis of such attributions is feeble. This complexity is compounded by the vast and simultaneous exchange of information occurring among individuals, between media and individuals, and across various media platforms. The sheer volume and interconnection of information make it challenging to comprehend fully, often leading individuals to abandon the pursuit of a comparatively convincing "version of the truth."

As new information is integrated into memory, the contents of memory expand, resulting in enriched and altered recollections. Individuals are frequently unaware of the profound effects that the intrusion of information can have on both personal and collective memory (Loftus, 1993). Such intrusions may precipitate existential and identity crises, wherein individuals experience a sense of emptiness, akin to a feeling of being ungrounded: LLMs as surrogate partners (Montag, Spapé, & Becker, 2025; Chandra et al., 2026). When fabricated information is more accessible than authentic information, the authenticity of reality becomes diminished. Ultimately, what we perceive and designate as 'reality' (Section 7) is also composed of conditional falsehoods. Upon further reflection, the indiscriminate use of NNs becomes detrimental when their outputs are presented as factual representations of real-life situations, prompting the question of what constitutes factual and real.

Individuals rely on agentic AI to do the work for them so to gain a competitive advantage, such as obtaining higher grades, enhancing their resumes, and achieving outcomes that would otherwise be beyond their capabilities. Consequently, on a *per capita* basis, artificial intelligence may elevate everyone's output to the extent that we ultimately find ourselves on equal footing once more. The underlying issue, however, is that by employing AI in this manner, we metaphorically don high heels to reach greater heights without actually growing taller. This creates a world of pretence that becomes progressively challenging to navigate from an epistemic perspective, as the mediated messages ultimately point to emptiness.

Humanity is in the process of constructing another filter bubble. LLMs generate a distorted worldview derived from biased training datasets, biased towards what the developers think the 'ground truth' is, compounded by hallucinations due to the irreducibility of the error term (Kalai et al., 2025; Chatterji et al., 2025; Chandra et al., 2026). Individuals subsequently disseminate these outputs as self-authored texts on the Internet, which are then absorbed by LLMs, thereby reinforcing the distorted worldviews initially formed with even more heavily biased training sets and unrecognised errors.

'When I see the generated result, then that is what I really thought.' LLMs are influential for the moral judgement of users, even when the advice provided lacks coherence. Despite being aware that they are interacting with a chatbot, users often underestimate the degree to which their judgements and decisions are affected (Chandra et al., 2026). While individuals maintain a high level of confidence in their authenticity, thinking they are true to themselves, their moral judgment and capacity for independent decision-making become compromised (Krügel, Ostermaier, & Uhl, 2023).

As an additional consequence, individuals who openly depend on systems like GPT or Gemini risk losing their authority. This loss of authority may occur because the NN is prone to hallucinations, exhibits bias, or lacks transparency regarding its sources, or because the system becomes so advanced that it surpasses the capabilities of its user. In either scenario, the user's credibility and authority are diminished. Currently, Microsoft incorporates advertisements into its 'new Bing,' resulting in scientific summaries interspersed with covert advertising from the highest bidders. This raises the question of why we should trust such content. We are approaching a communication crisis characterised by an inherent scepticism of mediated content, as it may be AI-generated and, therefore, inherently unreliable.

In moving away from predicting the next word toward actually solving complex multi-step problems, will the system show what type of problem solving was applied and tell us why it chose that approach: difference reduction, means-end analysis, forward-backward inference, analogical reasoning, or creative association? In reducing hallucinations (making things up) to a near-zero level, how do we know that what it now says is reliable? According to what world view or ontology, what science or maybe whose science does it cite? Science is a discussion that does not provide the ultimate answer. What if the user does not believe in science, but rather in religious truth or magic? Agentic behaviour is the ability for the AI to actually do things, like booking a flight, managing an inbox, or writing and executing code autonomously. What are the guardrails, how will the AI know what I want and allow, acting on my behalf? And who is watching over its shoulder without me knowing? Up till now, AI deployment has been encumbered by negligence and a lacks level of oversight.

In the remainder, we analyse the epistemic challenges posed by the reliance on AI systems for answering questions and evaluating outputs. We explore how this reliance can lead to violations of established communication maxims, such as Grice's Maxim of Quality, and legal principles, such as Lemoine's Maxim of Innocence. Our analysis highlights the authority fallacy inherent in trusting AI-generated content without understanding the underlying processes and biases.

We then move on to current integrations of semantic networks and ontologies, as well as logic-symbolic inference systems to address these challenges, offering a framework that accommodates diverse belief systems and epistemic perspectives. Additionally, we introduce Quantum Epistemics (Hoorn & Ho, 2026), a logic-symbolic inference system designed to update ontological classifications through comprehensive epistemic appraisal processes. Quantum Epistemics serves as a tool to navigate the complexities of AI-mediated communication, providing a means to critically assess AI-generated content and maintain a balanced perspective on reality perception. Through this exploration, we aim to foster the understanding of the implications of AI reliance and promote practices that prioritise transparency, proper citation, and the acknowledgment of generative and agentic AI's limitations.

## 2. The communication maxim

For a conversation to be successful, both parties must cooperate. The Maxim of Quality is one of the four cooperative principles Grice (1989) identified, centred on truthfulness and broken down into two specific rules: do not say what you believe to be false (i.e., do not lie) and do not say that for which you lack adequate evidence (i.e., do not state as fact what one is unsure of or has no evidence for).

The primary goal is to ensure that the information exchanged in a conversation is reliable. If we could not assume that agencies (whether human or artificial) were generally trying to tell the truth, communication would break down entirely. When a traveller asks a local for directions, they operate under the Maxim of Quality, the traveller may assume that the local will not intentionally point out the wrong way.

Irony and downright scams aside, humans who are aware of and obedient to the Maxim of Quality use what linguists call 'hedges' when they are not sure if they are meeting the requirement for evidence. Examples that most LLMs do not show, because they cannot do any epistemic appraisal themselves, are: "As far as I know...," "To our knowledge…," or "I'm not 100% sure, but I believe..." By using these phrases, speakers signal (Donath, 2007) to the listener that they are trying to respect the Maxim of Quality even though their evidence might be weak (also see Lee & Wagenmakers, 2013, p. 122).

The Maxim of Quality is the primary reason why people get frustrated with AI. When a model makes up a fact confidently, and states it with the supplementary rhetoric, it is failing the Maxim of Quality: it is saying something for which it lacks adequate evidence. In general, an LLM does not know what evidence it bases its claims on or the quality of that evidence (the sorcerer's apprentice). Therefore, the sorcerers are constantly trying to align, finetune, or 'post-train' the AI to better follow Grice's Maxim of Quality so that users can trust the information provided – in the eyes of the sorcerers, that is.

### *2.1 Not breaking the communication maxim*

Science is concerned with the formation of theory and the construction of knowledge, rather than the identity of the agency conducting the research. Focus on the latter is merely personal glorification. Consequently, scientists who are reluctant to engage with modern AI should view its emergence as a prompt to master the tool themselves; those who fail to adapt to this new technology may swiftly be left behind. Put differently, AI can be a great sparring partner as long as the user verifies and validates its suggestions.

A well-educated scientist can ensure that the information exchanged in a research article or conference contribution is reliable, truthful, evidenced, 'to the best of their knowledge.' When the use of AI is disclosed in the acknowledgements, the paper should not be dismissed as meaningless: particularly in the inductive phases of the research cycle, AI can generate alternative perspectives, which can be verified, validated, and may lead to new results with a probability attached to them. No matter what, epistemic responsibility for the work remains with the authors, while disclosure supports transparency, accountability, and reproducibility. Authors must thoroughly understand the AI's output in the same way they would any other raw material or data processed under their supervision or sampled during their research.

However, just as one should master arithmetic before using a pocket calculator, those who are not yet well-versed in academic rigour should not rely on AI until they have first learnt their craft and understood when it is appropriate to automate the more tedious aspects of research. Generative AI must not become a mere shortcut for students to avoid their fundamental coursework (Ju, 2023; Baidoo-Anu et al., 2024).

### *2.2 Breaking the communication maxim*

For non-experts of a discipline, LLMs cannot easily be fact-checked (Walther et al., 2009), because the training data are not openly available and users often miss the means to verify and validate an AI's claims. Irrespective, users generally assume the LLM is truthful (Grice), and they suspend disbelief until a counterfactual statement occurs, if users notice the contradiction at all (cf. truth trees, Beth, 1951). With LLMs, we should be moving towards an inverse conversational rule regarding the Maxim of Quality: the default assumption should be that generated information is unwarranted, even if it is not false.

The transition to digital communication has introduced significant opportunities for deception: in AI-mediated interactions, the central issues of honesty versus deception evoke theories from social media studies, such as Signalling Theory (Donath, 2007) and Warranting (e.g., Walther et al., 2009). AI-generated responses to human inquiries function as *conventional* signals (Donath, 2007); they are abstract and symbolic, consisting of digital words and images, and are therefore less reliable than *assessment* signals (ibid.), which are directly observable. *Conventional* signals require extensive fact-checking and warranting because they cue, for example, that an online-submitted assignment is authored by the student, which differs from witnessing the student physically write the assignment with pen and paper in the presence of the teacher (*assessment* signals).

Consequently, the AI or the individual utilising AI should provide evidence for each statement to enable the conversation partner to trust the artificial interlocutor. Ideally, this evidence should be non-mediated, implying that job interviews and student examinations will need to be conducted face-to-face, without digital assistance, to assess the candidate's genuine insights – personal insights, rather than mere knowledge that can be memorised or retrieved.

## 3. Frequentists administering the Turing test

Within the AI community, the conventional method for distinguishing between human and machine-generated work is to employ some version of a Turing test, comparing the AI to human performance (cf. emulation, singularity, super-human capabilities).[1] According to Turing (1950), if the results of the 'imitation game' are indistinguishable, the machine may be perceived as 'intelligent,' 'conscious,' 'sentient,' or exhibiting other human-like qualities.

Take notice, however, that Turing's perspective is phenomenological, suggesting that the machine is *psychologically* and *probabilistically* perceived in such a manner. Turing (1950) measures machine performance by an *operational* criterion: if a machine's, let's say, conversational behaviour is indistinguishable from a human's under specified conditions, we may choose to count that as intelligence for the purposes of the *test*.

By that pragmatic and empirical token, AI exhibiting human-like performance is a matter of definition or belief. For instance, Searle (1980) contends that the information processor in the Chinese Room is neither 'intelligent' nor 'conscious,' despite producing flawless output. On this view, a computer remains an extensive switch board, no matter what it says or does.

Turing's original setup of the test stipulated that one participant is a machine and the other a human, but modern versions keep the interrogator (participant) unaware of whom they are interacting with, in order to compare their different assessments. If we translate Turing's stance into test theory, then Turing-style tests state that if a machine's performance is indistinguishable from a human's (within some error bounds), then we are justified in counting it as 'thinking' or as

[1] Note, jackhammers also have super-human capabilities.

'intelligent' behaviour. There would be no good reason to deny it that description. Turing (1950) says that eventually, machines may eliminate observable differences (the test fails to reject the null hypothesis ($H_0$) at a chosen level of acceptability called 'significance'), hence we are justified in saying the machine thinks or is intelligent (as an aside, accepting backward inferencing like this would equally justify that humans are mere switch boards).

A Turing-style test becomes fallacious when behavioural indistinguishability is treated not as an operational criterion like Turing (1950) did, but as evidence of an unobserved inner property such as understanding, consciousness, or independent reasoning. The ELIZA bot in the 1960s was already seen as 'empathic' by its users (Weizenbaum, 1966). Doing so constitutes an inductive inference from observed behaviour to a mental-state attribution, probabilistic and defeasible, but an *affirmation of the consequent* nonetheless (Hoorn & Tuinhof, 2022): the cause is inferred from the effect without eliminating confounds.

| | |
|---|---|
| $A \rightarrow B$ | if intelligent, then it can reason |
| B | it can reason |
| $\therefore$ A | so it is intelligent (= $\perp$) |

In frequentist statistics, one can refute the $H_0$ or fail to do so, but not confirm the $H_A$. The Turing test conducted in a non-Bayesian testing style invites *affirming the consequent*, without indicating that evidence might be weak (Lee & Wagenmakers, 2013, p. 122), violating the Maxim of Quality once again.

Yet, we can push the argument even further. 'No differences detected' may merely indicate an impoverished signal sensitivity (e.g., not noticing LLM watermarks; statistical patterns meant to be imperceptible), or even personal bias, making a Type II error in wanting to believe that machines are a human's equals. In answering open-ended questions on their exams, sometimes students just write down something, not knowing what they are saying. The LLM also does not know what it says and writes equally tangential answers. Did the machine then pass the Turing test and should be considered intelligent? No noticeable difference.

That problem is most directly an appeal to ignorance (*argumentum ad ignorantiam*: Locke in Walton, 1992) or even an *absence-of-evidence* fallacy. No difference detected does not imply that no difference exists. The fallacious form is:

We have not detected that the conversation is machine-generated.
Therefore, the conversation is with a human being.

Or:

We have not detected a difference.
Therefore, the difference is not there.

Absence of evidence is treated as evidence of absence. However, while an AI's performance may be indistinguishable from that of a human being, this does not necessarily imply equivalence. The outcome is the same but the sources that produced it are entirely different. A car brings a person from A to B but that does not mean it is walking. Frequentist tests do not underscore equality or difference. They reject or fail to reject the $H_0$ under a predefined decision rule and rejection region, while controlling error rates. It is that mechanistic. In comparing human with AI output, frequentist methods may fail to reject the $H_0$ and, following Turing (1950), analysts may conclude that there is no discernible difference between the probability of AI-generated work and human-created work. However, if he could, Searle (1980) would argue that these are still two distinct distributions not originating from the same population; the measurement or the test statistic itself simply lacked the sensitivity to demonstrate the difference in intentionality or understanding between human and AI beyond reasonable doubt. Thus, it would be fallacious to conclude equivalence between AI and human performance.

## 4. The legal maxim

The situation is fundamentally altered when AI is responsible for fact-checking. We downloaded the full text of Turing's (1950) "Computing machinery and intelligence" and uploaded it to AskGPT for a plagiarism scan (29 March 2026). The result is found in Figure 1 (top panel), showing either that Turing copied 88% of his work from earlier authors ∨ that AskGPT missed by 12% that this was Turing's original ∨ that we could cheat unnoticed (no discernible difference) for one-eighth of the text ∨ that the missing 12% was made up of digital noise and formatting differences, not original writing. The epistemic issue that remains unresolved: which ∨ applies?

Plagiarism algorithms do not compare the paper as one giant block of text. They do not read or understand a text like humans do, but break a text down into overlapping strings of words (called *n*-grams: 5 to 7 words, so they use chunks of 5, 6, or 7 words, and why not 8 or 4?), while excluding stock phrases like "As we can see in the…" or "It is important to note…," as well as omitting stop words and function words such as 'the, and, an.' If a sentence breaks across two pages, the algorithm might read the page number or header as part of the sentence, breaking the match. Point is, in the text we uploaded, all the formatting was cleared out (no page numbers, headers, OCR misreading, etc.). If 12% was just noise, then did the similarity scanner tell us that 88% = 100% plagiarised? When we repeated the scan the following day, the level of plagiarism detected was 73% (Figure 1, bottom panel).

May the issue be database fragmentation? Online plagiarism checkers rarely match an upload against one single copy of a text: that one masterpiece that the human has read and that made such an impression that it became a classic in their heads. Instead, scanners find matches across thousands of different student papers, blogs of high-schoolers, and fragmented archives. Some LLMs have no access to, for instance, the IEEE database with peer-reviewed research articles. The machine has no quality standards. If the checker pieces together the match from multiple partial sources rather than one complete source, the transition points between those matched sources often slip through the cracks of the algorithm's matching criteria.

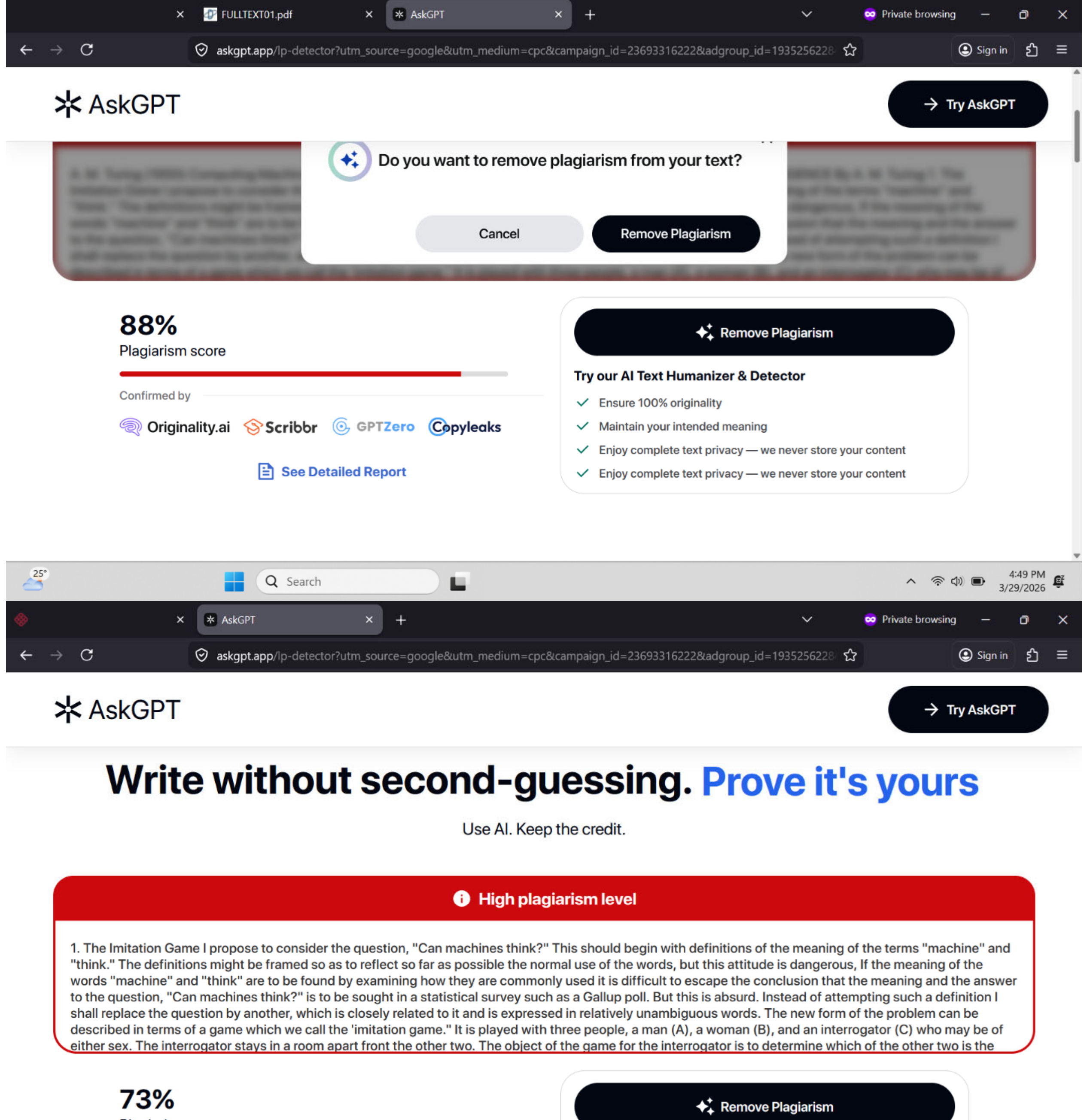

Figure 1. “Can machines think?” AskGPT. Prove it’s yours.

Probably, AskGPT uses an LLM to estimate plagiarism, rather than a traditional string-matching database like Turnitin or iThenticate, so the percentage is practically a hallucination. A stand-alone LLM does not itself search a database unless connected to retrieval tools; they predict the likelihood of text being generated by AI or being a known text based on statistical weights. They rarely output 100% for anything, because their mathematical make-up always leaves room for statistical variance (Kalai et al., 2025; Chatterji et al., 2025): write without second guessing?

Even with perfect formatting, the algorithm is mathematically designed to ignore certain words, penalise punctuation or spelling variations (i.e., Turing's British English matched against US databases), and process text in rigid chunks with varying numbers of words per chunk. Hitting exactly 100% is actually quite rare for long documents (Turing's is about 11 to 12,000 words), even when the entire text is copied in, as we did.

Why a 15% overnight drop to 73%? Depending on server traffic the very next day, the API might have only processed the first 5,000 words before timing out, or it might have summarised the text differently before analysing it. If the server was busier the second day, it might have processed less of the text, changing the score. Some GPT-wrapper tools try to combine AI with live web searches. If it broke Turing's paper into different chunks and Googled them, the results can change daily. Possible explanations include changes in retrieval, chunking, pre-processing, stochastic model behaviour, database availability, or server-side limits. Without access to AskGPT's methods, the cause cannot be determined. It is known that search APIs frequently time out, encounter CAPTCHAs, Completely Automated Public Turing tests to tell Computers and Humans Apart (*sic*), or return different indexed results from day to day. If the tool's search API failed to fetch results for a number of those chunks on the second day, the plagiarism score would plummet. That the plagiarism score is capable of changing at all for identical text shows that AskGPT is instable, hence, unreliable, its results non-reproducible, or underspecified. The change from 88% to 73% shows that the score is not reproducible under the observed conditions and therefore cannot be treated as a stable measurement without disclosure of the method, database, pre-processing, and error rates. It recognises that the text is known, which is why the number is high, but the specific percentages it gives are mathematically meaningless guesses generated by an AI attempting to sound authoritative: "Use AI. Keep the credit."

If a student submits original work, but an AI-enhanced plagiarism scanner determines that the similarity rate with AI-generated, which is copied, content is excessively high (73%, 88%, 100%?), the student unexpectedly bears the burden of proving that s/he did not engage in cheating. In the absence of contrary evidence, the student will be presumed guilty.

This assumption contravenes a foundational principle of Western legal systems, which can be traced back to the 13th-century Cardinal Jean Lemoine. He articulated the maxim *item quilbet presumitur innocens nisi probetur nocens*, meaning that an individual is presumed innocent until proven guilty (Ullmann, 1950; Pennington, 2003), after which the Inquisition would torture people to a coerced confession – *ad baculum*.

John Locke introduced the term *argumentum ad ignorantiam*, referring to an appeal to ignorance, which denotes the lack of counter-evidence. Although the formal logic may be structurally sound and the accused empirically guilty (Walton, 1992), within the realm of informal logic, an appeal to ignorance constitutes a fallacy. However, such appeal is considered justified because the prosecutor bears the burden of proof; failing to uphold this principle would compromise notions of honesty and fairness, potentially resulting in widespread accusations and social instability. This serves as a cautionary note regarding the reliance on contemporary plagiarism detection tools, because to tell computers and humans apart, they do not handle their *t*-tests right (if at all), neither according to Fisher (1932), Neyman-Pearson (1933), Lindquist (1940), nor to Bayes (Jeffreys, 1961, p. 283).

Where things go wrong is that presumption of innocence does not mean that what has not been proven false must be considered true; it means something far more procedural: the burden of proof lies with the accuser; absent sufficient evidence, no penalty should be imposed. In court, a judge does not decide for innocence when s/he rules 'not guilty.' The judge acknowledges that guilt has not been demonstrated to the required standard (beyond reasonable doubt: $p < \alpha$).

The presumption of innocence is not an inference from "not proven guilty" to "proven innocent." It is a burden-of-proof rule: unless misconduct is established by adequate evidence, the accused should not be sanctioned. But this is not what is happening. In the context of plagiarism detection, the reverse is applied to the detriment of the accused, whereby the student is presumed guilty until proven otherwise. This represents a simplistic bipolar epistemic framework, where truth is equated to the negation of falsehood, i.e., true = 1 – false. In contrast, a unipolar epistemic perspective allows for the coexistence and mutual influence of truth and falsehood, without necessarily summing to one; they interact in dynamically varying proportions, akin to the adjustments made using an audio equaliser. Yet an alternative epistemic approach suggests that truth and untruth form a bi-dimensional unipolar scale, which is related to another bi-dimensional unipolar scale of false and not-false. These scales exist in a superposed state, creating ambiguity regarding the final verdict while the judicial process remains ongoing (Hoorn & Ho, 2026; Ho & Hoorn, 2022).

## 5. The reverse Turing test

No detectable difference from AI should not be treated as evidence of AI authorship, and AI outputs and checkers should not be trusted without warrant. In the context of a reversed Turing test, the artificial intelligence system (e.g., the plagiarism scanner) determines whether it is interacting with another computer or a human being (Feigenbaum, 2003); whether the work presented is generated by AI, created by humans, or a combination of both. While this approach may be persuasive, it shifts the syllogism of *argumentum ab auctoritate* (Fellmeth & Horwitz, 2009) from a human authority to an 'expert' AI (e.g., AskGPT), thereby deriving the argument from the authority of a machine and its training base.

The concept of expertise is crucial. Although an appeal to authority is generally considered fallacious, it may be deemed non-fallacious if the authority in question is, to the best of our knowledge, the most knowledgeable source available. However, this always remains uncertain. Regardless of the epistemic framework or epistemology in use, reliance on authority is fundamentally a matter of belief once more, induction, probability, and likelihood. Definitive proof can only be given in closed systems, such as classical logic and mathematics, and not in contexts related to open systems or those susceptible to information entropy. A complete fallacy occurs when belief is placed in a non-authoritative source, one that lacks broad consensus, such as the next version of yet another LLM release.

If the burden of proof is not assigned to the system, reversing the Turing test, as exemplified by the plagiarism scanner (Figure 1), may easily constitute an *authority fallacy*, causing a further infringement upon the legal principle of presumption of innocence. Using an instrument is not automatically an appeal-to-authority in a fallacious manner. It becomes so, when users treat the instrument's output as decisive without validation, error rates, calibration, source evidence, or procedural safeguards. The fallacy arises not so much from using a detector, but from treating an obscure, unvalidated score as authoritative evidence. For instance, the system should not merely indicate superficial similarity but should ascertain the origin of the copied material; as in court: exhibit A, B, C to support its case. Expert authority can be legitimate when properly warranted.

The university basically says to its students that we suspect you used AI for your assignment and if so, your uploaded text will look AI-like. Indeed, the plagiarism scanner says it is AI-like (88%). Therefore, we know that your text was largely AI-generated:

A: if your text is AI-generated,
B: it resembles AI output that is not detectably different from AI.

Thus,

| | |
|---|---|
| $A \rightarrow B$ | if AI-generated, then AI-like |
| B | AI-like (88%) |
| $\therefore$ A | so it was mainly AI-generated (= $\perp$) |

The defect lies in the unsupported premise that human work must differ detectably from AI output: if AI wrote it, it may resemble AI-output. I see it resembles AI-output, so AI must have written it. This reasoning *affirms the consequent* and ignores base rates, genre conventions, and missed signals (Type II errors). Human academic prose may resemble AI output because AI is trained on human academic prose, because students learn standard academic phrasing, because foreigners repeat language patterns unlike native speakers do, and because genre conventions constrain wording.

The process that is happening is this: a Type II statistical error occurs: the test is insensitive or lacks power and misses the difference. A person interprets "not detected" as "not present," which is a logical fallacy. Then the wrong policy of the institution is to treat a high (where's the cut-off point?) but uncertain score as disciplinary evidence.

In a fair plagiarism or AI-detection context, a frequentist tests should specify the $H_0$, the $H_A$, select a test statistic (e.g., 2-tailed independent samples *t*), and a decision rule (e.g., $\alpha = .05$). The ethically appropriate null hypothesis would assume not guilty until proven:

$H_0$: the student wrote this work (no misconduct, no AI interference)
$H_A$: the work is AI-generated or plagiarised

Then if a Type I error is committed, concluding "AI/plagiarised" when the student actually wrote the piece herself, a false accusation is returned. In the case of a Type II error, concluding "student wrote it" when it actually was AI/plagiarised, a case of fraudulent behaviour slipped through the net. Currently, we see institutions flip the approach and implicitly assume:

$H_0$: the work is AI-generated or plagiarised
$H_A$: the work was written by the student

In this setup, the institution effectively treats the student as "guilty until proven innocent," whereas the AI checker should be treated as such! Failure to demonstrate difference from AI is treated as confirmation of the null: "Look how strict we are; we eagerly conclude for AI/plagiarism." This is a problem of hypothesis specification and presumption, which academia apparently are not good at!

The AskGPT scanner output in Figure 1 actively labelled the uploaded article as AI/plagiarised. Under the natural, fair $H_0$ ("student wrote it"), this would be a Type I error: rejecting $H_0$ (student wrote it) when $H_0$ should be maintained, falsely accusing the student, here, the student named Alan Turing, of self-plagiarism. The assertion that one's work must differ from AI-generated content to be considered original implies that any lack of difference suggests the work is not one's own. This reasoning leads to the presumption that a student is guilty of plagiarism until demonstrated otherwise. In sum, the approach comes from a mis-specified null hypothesis. Operationally, the test behaves as if:

$H_0$: not the student's work (AI/plagiarism)
$H_A$: student's own work

In the music industry, individuals are free to sample as much music as they wish, provided they acknowledge the source (and, in protectionist societies, compensate accordingly). This approach aligns with good open-source practices, fostering creativity and innovation, and contributing to a continuous body of scientific work where it is evident who is standing on "ye shoulders of Giants" (Newton, 5 Feb. 1676, in a condescending letter to Robert Hooke).

The question of how small a portion of sampling remains acceptable is complex. Even if a piece of music is deconstructed into individual notes, the composer may still be considered to be copying from the original work, despite not replicating the exact arrangements. This principle similarly applies to language. In scientific discourse, clarity necessitates that each technical term possesses one singular, precise meaning. This requirement extends to the use of equations and formulae. Employing synonyms indiscriminately or paraphrasing equations to circumvent plagiarism can lead to confusion and may constitute academic misconduct.

In Figure 1, users are invited to 'humanise' their texts and by humanising the AI-generated content, the student suggests that:

| | |
|---|---|
| $A \rightarrow B$ | if human-generated, then human-like |
| $B$ | human-like (88%) |
| $\therefore A$ | then my text is human-generated (= $\perp$) |

'Humanising' basically means writing a layman's version of an academic text while losing nuance and precision. For example, the thing may change the phrase "making validated claims" into "using verifiable facts," which epistemically, is a completely different outlook on the very existence of objectivity, on what scientists are producing in their labs, on the difference between validation and verification, and hence, on the trustworthiness and interpretability of results. How about non-verifiable facts? Are they still facts? How does one know? Or is the rewrite a pleonasm? All of this is not linguistic nit-picking; this particular rephrasing by AI articulates an almost opposite view on knowledge itself (i.e., the standard view). For the layman, that does not make a difference (because it is the standard view). In class, we try to optimise the student's sensitivity to the meaning of the words they choose, but at face value, AI output sounds good, we have to admit.

Nonetheless, the current default in academia is suspicion of the student, not of the AI, and the burden is on the student to show enough difference. Such misuse of test theory comes from flawed logics. The invalid logical inference moves from "genuine work tends to differ from AI" to "if there is no detected difference from AI, it is not genuine work." This is all too strange since the bulk of the AI's training base is the very academic work the student should be different from and so, s/he is forced to become less academic. Together, these flaws amount to the violation of the legal maxim (Section 4), maintaining that "one is presumed guilty until proven innocent."

## 6. Latest approaches

An artificial advisor must adhere to three constraints: 1) evidential, making validated claims (science) or admitting uncertainty (blogs), integrating extra-logical world knowledge or ontologies in a disciplined way, 2) relational (prioritising accuracy over user rapport), and 3) its role (avoiding perceived authority and ensuring paths for human escalation).

In our Research Laboratory for Advanced Social Robotics (EARNEST), rather than relying on unconstrained generation, the LLMs we employ in critical situations such as depression treatment are safeguarded by curated knowledge sources, rule repositories, question-answer templates, and structured guidelines (e.g., professional depression protocols). Our basic setup is that inferential AI produces the prompt, the LLM formulates a range of naturally sounding variations, after which the inferential AI selects what is appropriate. In doing so, the LLM becomes an interface widget with an overlay of expert-system guardrails. Varieties of formulations, which could run into the millions, go into a repository from which in the future the inferential AI can draw, after which the LLM can be depreciated, so that risk is reduced in critical application environments (more controllable, better auditable), the entire system being implementable on edge devices (i.e., the robot itself) without third-party surveillance (Hoorn, 2018), while saving resources (no bulky data required). In our approach, the deductive core is not performed by the LLM itself, but by an external symbolic reasoner with the LLM acting as parser, planner, or explainer.

Pure NNs are poor at reasoning because they are built on association (i.e., correlation), variation, and probability (Kalai et al., 2025; Jiang et al., 2025). This is why serendipitously, NNs can do categorical cross-overs that humans recognise as creative (Hoorn, 2022; 2023). Currently, the strongest systems are hybrid neuro-symbolic or agentic systems in which the LLM handles formalisation, search guidance, repair, or explanation, while the deductive capability comes from an external prover or proof assistant (e.g., Callewaert, Vandevelde, & Vennekens, 2025/2026; Cao et al. 2025; Liang et al., 2025).

A logic-LM explicitly uses the LLM to translate natural language into a symbolic formulation and then runs a deterministic symbolic solver; for instance, Logical Inference via Neurosymbolic Computation (LINC) does that with a first-order logic prover, the formal deduction being offloaded (Olausson et al., 2023). The justificatory authority is located within the broader human-LLM-prover constellation, where validity is distributed over LLM training base and external ontology (e.g., Guo et al., 2025), formal verification resides in the symbolic component, and the human *must* understand and *is responsible* for both (cf. Stoyanov, 2026). Frontier reasoning models improve empirical accuracy, but we have not seen evidence yet that establishes native formal soundness, proof faithfulness, or machine-checkable validity for standalone LLM outputs; *benchmark success should not be conflated with trustworthy theorem proving* in the style of Beth (1951) (e.g., Hoorn & Tuinhof, 2022). As far as we can see, there is no evidence that LLMs by themselves can perform Beth-style tableaux or formal proof calculi (Abzianidze, 2023; Lin et al., 2025; Liu et al., 2025; Pan et al., 2023; Zhou et al., 2025; Xu et al., 2025).

Several further claims in the field remain plausible but unproven, including the possibility that agentic orchestration will generalise reliably beyond benchmark settings, that auto-formalisation will become dependable enough for unsupervised use, and that translation bottlenecks in hybrid systems can be largely overcome. Still unproven are stronger theses, such as whether LLMs possess internal representations genuinely isomorphic to formal proof systems, whether scaling alone can yield trustworthy deduction, or whether current agentic systems are robust enough for safety-critical deployment.

## 7. An epistemics of the virtual: meta-inference about the knower

In the previous, we saw that AI may hallucinate or misuse logic (also see Zheng et al., 2025) but, in this section, we argue that every act of perception, classification, and judgment is conditioned by an observer framework, whether organic or synthetic.

> …we should *not* think of delusional spiraling as a symptom of lazy, irrational, or fallacious thinking from users, or as the result of insufficient epistemic vigilance on the part of users. (Chandra et al., 2026, p. 6)

If so, then in the Research Laboratory for Advanced Social Robotics in the Hong Kong SAR, we are developing an implementation of Epistemics of the Virtual (Hoorn, 2012) to describe belief formation (Chandra et al., 2026, p. 6), stating that the physical world can be assessed by sensors, albeit in a limited and potentially biased manner. Consequently, what is perceived as data that represent 'reality' is inherently conceptual and subject to change.

Some scientific entities such as dark matter or white matter are known through instrument-mediated observation, indirect inference, or theory-laden measurement rather than unaided perception. Quite a number of scientists believe in the possibility of creating a wormhole on a quantum chip to teleport a subatomic particle (Jafferis et al., 2022). Conversely, some critics accuse these people of perpetuating a hoax (i.e., Woit, 2022). Similarly, non-experts may be inclined to accept the highly realistic outputs of the GPT series as truth, despite these being mediated summaries derived from unknown sources. The notions of Earth being flat, spherical, or an irregular ellipsoid represent three distinct ontologies that should be managed with equal proficiency, reflecting different worldviews or perspectives. Our Quantum Epistemic system (Hoorn & Ho, 2026) can transition from a 'flat' to an 'irregular ellipsoid' model by broadening the criteria for acceptance and tolerance for deviation, or conversely, by narrowing the parameters of what is deemed acceptable, including metaphors such as Earth being a "blue marble."

That we should accept flat, spherical, and an irregular-ellipsoid Earth as distinct ontologies is because a system may need to model different users' beliefs, including an Alzheimer's delusion that the nurse is a ghost and the deceased husband alive. Certainly, as an epistemic claim, that may become dangerous (think of medication refusal, faith healing), because it may imply that empirically unvalidated and empirically warranted ontologies deserve equal standing. Thus, different ontologies should be representable with equal technical accuracy, but a developer could choose to not treat them as equally warranted. Trustworthy AI, like a medical doctor, takes the *ethical* stance that to 'the best of its knowledge,' it offers what is 'best for its user,' although all of this, indeed, should be taken under epistemic relativism: not everyone shares the same list of beliefs and arguments that convinces them and there is no way of telling who is right or wrong. We can merely make a case.

Quantum Epistemics is an observer-effect framework (Hoorn & Ho, 2026), extending the approaches discussed in Section 6 by relocating the epistemic problem from the level of output reliability alone to the more fundamental level of observation, classification, interpretation, and beliefs. The preceding discussion has mainly focused on the limitations of stand-alone LLMs, the need for symbolic verification (e.g., theorem provers), the role of ontologies, and the importance of distinguishing verification, validation, and explanation. By contrast, Quantum Epistemics argues that these issues arise within a prior layer in which the observer is already inseparable from the epistemic process. On this view, classification is not a neutral mapping from world to representation, nor is judgement a simple read-out from data, but rather the outcome of an interaction between sensory information and observer states, where prior beliefs (cf. Bayes), affective dispositions, and degrees of openness or disbelief modulate what is perceived and how it is classified (Hoorn & Ho, 2026). Thus, we shift attention away from the trustworthiness of AI outputs to the trustworthiness of the conditions under which inputs are rendered meaningful.

Much of the preceding analysis has treated epistemic improvement in terms of external warrant. In this section, we claim that trustworthy AIs are systems in which the LLM is constrained by curated sources, ontologies, rule repositories, external solvers, proof assistants, or verifier-guided loops. In each of these cases, authority is displaced away from unconstrained token prediction and relocated to symbolic machinery that can provide formal guarantees or at least structured grounds for assessment. Quantum Epistemics does not reject this trajectory; it broadens it by arguing that even formally checked systems presuppose a stage at which information has already been encoded, weighted, segmented, and rendered classifiable by an observer-dependent process. In that sense, accounting for the observer effect adds a pre-verification layer to the epistemic analysis. Formal verification remains indispensable for theorem proving and critical applications, but Quantum Epistemics suggests that verification alone cannot cover the question of trust, because the object being verified has already been shaped by the interaction between observer and observed (Hoorn & Ho, 2026). As an overly simple example: American databases checking Turing's British English.

A central contribution of the observer framework is that it provides a formal account of subjectivity rather than merely identifying subjectivity as a general concern. In the foregoing sections, subjectivity appeared in various forms: user naivety, authority fallacies, hidden assumptions in plagiarism detection, differing ontologies, and the influence of worldview on what counts as evidence. Quantum Epistemics goes further by modelling such variation mathematically, dealing with ambiguity and vagueness in the clearest of ways. Sensory information is represented as a structured state space that evolves through interaction with observer states before any final decision is reached. This pre-decision evolution is described with the Lindblad master equation, while classification itself is modelled through adaptive Positive Operator-Valued Measures (POVMs), allowing the observer's stance to modulate the eventual outcome distribution (Hoorn & Ho, 2026). Differences in judgement are no longer treated solely as noise, individual bias, or technical defect, but as formal consequences of the measurement process itself.

This changes the interpretation of ontology or 'ground truth' within AI systems. Earlier in our paper, ontologies and semantic networks were discussed as explicit symbolic structures that can support grounding, consistency, traceability, and formal reasoning. Within Quantum Epistemics, however, categories are not merely external structures to which data are matched. They are also functions of the observer's resolution, available feature distinctions, truth-value granularity, and tolerance for ambiguity. In other words, an ontology is not only what a system knows; it is also the way in which a system is able to classify what is in the world. In broadening our earlier argument, ontological design is not purely technical, because decisions about what counts as a feature, what counts as an attribute, and how uncertainty is represented already embed assumptions about the observer's epistemic stance. This is particularly relevant to hybrid AI systems, because even when formal reasoning is delegated to symbolic engines, the **categories supplied to those engines may already reflect observer-dependent commitments**. Also, a Beth tableau relies on truth conditionals to decide for contradiction and the closing of inference branches.

Quantum Epistemics enriches the distinction between verification, validation, and explanation. A formally verified proof may remain opaque, and a fluent explanation may aid understanding without itself providing proof. The observer-effect account adds another dimension by showing that explanation is not only distinct from proof, but may itself be observer-modulated.

If classification outcomes depend partly on the observer's cognitive and affective state, then explaining those outcomes cannot be limited to displaying derivations, proof objects, or natural-language rationales. It also requires disclosing the observer parameters that shaped the result, such as how similarity and dissimilarity were weighted (cosine, Hammings, Euclidian, dot-product, Tversky-index?), how absent features were interpreted, and how much uncertainty or incompleteness was tolerated (i.e., hyperplane, softmax). To have trustworthy AI, interpretability is not exhausted by transparency of the symbolic backend. A system may be formally sound while remaining opaque with respect to the observer conditions that govern how it frames and measures the world in the first place. After all, commercial AI developers sit on their training data and keep them secret.

Quantum Epistemics is especially relevant to the earlier critique of plagiarism detection and reverse Turing tests. We criticised AI plagiarism scanners for mis-specified null hypotheses, inverse reasoning, authority fallacies, and violations of the legal maxim of innocence. Within the account of the observer-effect, such systems are not neutral detectors but observer-structured measurement devices. They embody specific strategies of similarity and dissimilarity, encode particular thresholds for ambiguity, and operationalise hidden assumptions about what authorship looks like and how deviation from expected linguistic patterns should be interpreted. In this sense, the plagiarism detector is not merely a tool that reads off a pre-existing fact from the text; it is itself an observer with a built-in epistemic bias. What appears as a factual classification, such as "AI-generated" or "plagiarised," is therefore the outcome of a measurement regime whose assumptions are already shaping the object under judgement. The problem is not only that these systems infer invalidly, but also that they conceal the observer-dependent setup through which they produce their classifications.

One can analyse propositions, hypotheses, proofs, and inferential errors, overlooking the pre-propositional stage of reasoning. Quantum Epistemics directs attention to what happens before propositions are even formulated: the encoding of sensory input, the assignment of truth-values, the selection and weighting of features, and the adaptive formation of category structures. Hence, failures attributed to "reasoning" may have originated earlier (cf. error cascading), in how the world has already been dissected for reasoning. The antecedent problem may be one of observer-conditioned perception and categorisation (pre- and post-training the AI) rather than an LLMs deductive failure by itself. A logical inference may be formally impeccable (verified as 'true') while still resting on contestable inputs ('this bicycle is my property because it is mine'), because those inputs have already been transformed by interpretive routines and observer states prior to any explicit proposition being tested, read, or empirically validated.

User uncertainty transpires with weak evidence, hallucination, benchmark failure, false positives and negatives, and the absence of formal soundness. Yet, in a quantum-probabilistic formulation, uncertainty is not merely a deficit of information or an error term to be reduced. It can also be the coexistence of several possibilities, modulated by the observer's own state and by the measurement procedure adopted. An ardent believer in how things are and are supposed to be (e.g., 'I saw AI show signs of sentience') tends to impose narrower boundaries and stricter classification, treating the absence of evidence ('it's but a switch board') as a stronger indication of non-membership ('that does not count'), whereas a more self-sceptical observer allows broader interpretive spread and more tolerance for incomplete information (Hoorn & Ho, 2026). Such variation represents different epistemic views, not defects, which affect how ambiguity is resolved. Like this, uncertainty goes further than just making a Type I or Type II error (Section 3).

It is often assumed that the strongest source of epistemic warrant remains formal verification. With its observer-effect approach, Quantum Epistemics does not undermine that conclusion, but it does alter its framing. If the observer must be treated as part of the epistemic system, then formal proof checking secures validity only within a particular representational and observational setting. This means that trustworthy theorem proving still depends on kernels, proof assistants, or external checkers, but that the surrounding questions of relevance, intelligibility, task framing, and responsibility cannot be addressed by proof acceptance alone. The quantum modelling of the observer effect, therefore, reinforces the claim that epistemic authority in AI is distributed and layered, while adding that one of those layers is the observing system through which problems are perceived, encoded, and prepared for proof and evidence in the first place.

Note that Quantum Epistemics is not an engineering replacement of symbolic reasoners, proof assistants, or ontology-backed systems but a meta-epistemic framework within which those systems can be interpreted. It asks for the hidden assumptions of verifiers and neuro-symbolic systems, wants to know what kind of observer is implied by a given ontology, by a given similarity measure, by a given proof task, or by a given detection pipeline. In that respect, it changes the discussion from one of technical composition to one of epistemic reflexivity.

Many developers ask how AI systems can be made more reliable by constraining LLMs with external procedures of justification, including human feedback. Quantum Epistemics asks an earlier question: what counts as evidence, classification, and even observation once the observer is formally included in the system? 'Truth' as mainly a property of propositions and proofs now becomes a 'truth judgement' emerging from relations among information, observer state, category structure, and measurement procedure. We avow that trustworthy AI requires not only better reasoning engines but also an explicit model of the observer conditions under which reasoning becomes possible or is happening.

Associative language models are insufficient for trustworthy reasoning. External symbolic verification, ontologies, and human oversight are needed to secure warrant. Quantum Epistemics adds that warrants are established within a prior interpretive regime in which observer states, measurement strategies, and category structures shape what can appear as evidence at all. A software architecture of trustworthy AI would therefore require at least four interconnected levels: an *observer-aware epistemic layer*, **the overwatch**, that models stance, salience, ambiguity tolerance, and interpretive bias; an *ontological layer*, **the erudite**, that structures domain knowledge and admissible concepts; a *symbolic verification layer*, **the logician**, that provides formal guarantees where possible; and a *linguistic interface layer*, **the interlocutor**, that enables communication, explanation, and interaction. The reliability of AI cannot be reduced to reasoning correctness alone, but depends also on the observer conditions through which the world is rendered into something that agencies, humans and AIs, reason about.

## 8. Conclusions/Discussion

The reliance on AI to answer our questions and evaluate outputs presents significant epistemic challenges that must be addressed to maintain the integrity of knowledge and communication, maybe even of science itself. The use of LLMs like GPT, Gemini, Grok, can lead to a violation of Grice's Maxim of Quality, as these systems often produce untransparent outputs, unverified and unvalidated. Users might unknowingly accept AI-generated content as factual, leading to a distorted perception of the world about us. The authority fallacy is also at play, as users may attribute undue credibility to AI outputs without understanding the underlying processes or biases of the data and algorithms involved (cf. Lam et al., 2024). This reliance on AI can result in errors, where false information is accepted as true or validated information is rejected as false due to the absence of clear assessment signals (Donath, 2007). To mitigate these issues, it is crucial to develop AI systems that can explain their decision-making processes and incorporate logic-symbolic inference to handle diverse belief systems and epistemic frameworks.

Outputs from stand-alone LLMs do not provide formal rigour, and current LLMs are not reliable substitutes for symbolic proof systems in tasks requiring formal evaluation. If LLM-based theorem proving happens, it is not one unified phenomenon, and epistemic authority in these systems does not belong simply to the language model. Instead, authority is distributed across different components and paradigms like formal verifiers, symbolic solvers, retrieval systems, agentic orchestration pipelines, explanation layers, and human or institutional oversight, which is very important. Users should stop collapsing very different systems into a single category; the source of trust differs sharply across them.

We have argued that formal verification remains one of the strongest bases for epistemic warrant: a proof checked by a formal system has a stronger claim to authority than an informal but persuasive natural-language explanation. We urge that explanation, verification, and validity must be separated: fluent "thoughts" (which are correlational associations) or proof-like narratives may help search or improve usability, but they **do not themselves make a theorem true**. In recent frontier systems, successful proving often comes from complex workflows rather than one model's intrinsic processing, so responsibility and authority are distributed. Overall, our work underscores that **assessing these systems requires a taxonomy of paradigms and a normative framework for verification, accountability, and epistemic credit**.

Overall, we identified that stand-alone LLMs are not yet trustworthy theorem provers; hybrid systems outperform stand-alone models on logical reasoning tasks; verifier-loop architectures can provide machine-checked proofs; and no evidence shows that standalone LLMs natively implement Beth-style tableaux (e.g., Cailler et al., 2022) or comparable formal proof calculi.

Furthermore, the indiscriminate use of AI in generating and evaluating content can lead to a communication crisis characterised by disbelief and mistrust. As AI systems become more advanced, they risk surpassing human capabilities, leading to a loss of authority and credibility for users who depend on them. This situation is exacerbated by the integration of biased training datasets and the potential for AI to produce hallucinations, further distorting the information available to the user. To navigate this complex epistemic environment, it is essential to prioritise transparency, proper citation, and the acknowledgment of AI's limitations.

The observer-effect framework of Quantum Epistemics adds a deeper epistemic layer to the approaches discussed above. Whereas our present analysis emphasises that stand-alone LLMs lack formal warrant and require ontologies, symbolic reasoners, proof assistants, or verifier-guided architectures, Quantum Epistemics argues that an even earlier problem must be addressed: observation and classification, data and algorithms, are themselves not neutral. Rather than treating AI error only as a matter of hallucination, weak inference, or failed verification, in need of tweaking bias and feeding it more training data, it formalises how sensory information is already modulated by the observer's prior beliefs, affective state, interpretive stance, and tolerance for ambiguity. In this view, subjectivity may be a practical nuisance but it also is a structural feature of epistemic processing.

This does not replace neuro-symbolic or verifier-centric AI. Formal verification remains a strong source of warrant in theorem proving and other critical domains. However, Quantum Epistemics changes how such systems should be understood. It suggests that even formally verified outputs presuppose observer-dependent choices about meaning: feature salience, category construction, similarity and dissimilarity, and the handling of missing information; in all, the reference to the empirical world. Trustworthy AIs are systems that can reason correctly under external symbolic control and whose observer assumptions are made explicit. On that note, we would welcome an epistemological, not purely engineering, shift in the current AI discussion.

## Acknowledgements

This study is funded by the Research Grants Council (project code: T43-518/24-N) under the University Grants Committee, Hong Kong Special Administrative Region Government. We thank Becky Bik Kei Chan for the literature search.